\documentstyle[12pt]{article}
\begin{document}
{\hfill PUTP-95-22 (revised version for PUTP-94-12)}
\vspace{10mm}
\begin{center}
{\bf PSEUDUSCALAR HEAVY QUARKONIUM DECAYS WITH BOTH RELATIVISTIC AND QCD
RADIATIVE CORRECTIONS}
\vspace{10mm}\\
{Kuang-Ta Chao$^{1,2}$~~~Han-Wen Huang$^{2}$~~~ Jing-Hua Liu$^{2}$~~~
Jian Tang{2}}\\ 
\vspace{4mm}
1.{\it CCAST (World Laboratory),~Beijing 100080, P.R.China}\\
2.{\it Department of Physics, ~Peking University,
	 ~Beijing 100871, P.R.China}
\end{center}         
\vspace{8mm} 

\begin{abstract}
We estimate the decay rates of $\eta_c\rightarrow 2\gamma$, 
$\eta_c'\rightarrow 2\gamma$, and $J/\psi\rightarrow e^+ e^-$, 
$\psi^\prime\rightarrow e^+e^-$, 
by taking into account both 
relativistic and QCD radiative corrections. The decay amplitudes 
are derived in the Bethe-Salpeter formalism.  
The Bethe-Salpeter equation with a QCD-inspired 
interquark potential are used to 
calculate the wave functions and decay widths for these
$c\bar{c}$ states. We find that the relativistic correction
to the ratio $R\equiv \Gamma (\eta_c \rightarrow 2\gamma)/ 
\Gamma (J/ \psi \rightarrow e^+ e^-)$ is negative and 
tends to compensate the positive contribution from
the QCD radiative correction. Our estimate gives $\Gamma(\eta_c
\rightarrow 2\gamma)=(6-7) ~keV$ and $\Gamma(\eta_c^\prime
\rightarrow 2\gamma)=2 ~keV$, which are smaller than their 
nonrelativistic values. 
The hadronic widths $\Gamma(\eta_c
\rightarrow 2g)=(17-23) ~MeV$ and $\Gamma(\eta_c^\prime
\rightarrow 2g)=(5-7)~MeV$ are then indicated accordingly 
to the first order QCD radiative correction,
if $\alpha_s(m_c)=0.26-0.29$. The decay widths for
$b\bar b$ states are also estimated.
We show that when making the assmption that the quarks are on
their mass shells our expressions for the decay widths will become 
identical with that in the NRQCD
theory to the next to leading order of $v^2$ and $\alpha_s$. 
\end{abstract}
\vspace{10mm}

Charmonium physics is in the boundary domain between perturbative
and nonperturbative QCD. Charmonium decays may pvovide useful
information on understanding the nature of interquark forces and 
decay mechanisms. 
Both QCD radiative corrections and relativistic corrections are
important for charmonium decays, because for charmonium the strong 
coupling constant $\alpha_s(m_c)\approx 0.3$ [defined
in the $\overline {MS}$ scheme (the modified minimal subtraction
scheme)] and  the velocity squared of the quark in the meson rest frame
$v^2\approx 0.3$, both are not small.
Decay rates of heavy quarkonia 
in the nonrelativistic limit   
with QCD radiative corrections have been studied (see, e.g., refs.[1,2,3]). 
However, the decay rates of many processes are subject to substantial 
relativictic corrections. 
In the present paper, we will
investigate relativistic corrections to the pseudoscalar quarkonium 
decays such as  
$\eta_c \rightarrow 2\gamma$ and $\eta_c^\prime \rightarrow 2\gamma$,
and give an estimate of their widths
by taking into account both 
relativistic and $QCD$ radiative corrections. (For a brief report on
this result, see also ref.[4].)
For comparison we will also study the leptonic decays of the vector 
charmonium such as $J/\psi\rightarrow e^+ e^-$ and $\psi^\prime
\rightarrow e^+e^-$.

These pseudoscalar charmonium decays are interesting. Experimentally,
the branching ratio of $\eta_c \rightarrow 2\gamma$  may provide
an independent determination of $\alpha_s$ at the charm quark mass, but
the measured $\Gamma (\eta_c\rightarrow 2\gamma)$ ranges from 6 keV
to 28 keV [5], and the measured $\eta_c$ total width is also uncertain. 
As for the $\eta_c^\prime$, its exsistence needs to be confirmed, and
its two gamma decay mode is being searched for by the E835 experiment
at the Fermi Lab $p\bar p$ collider,
and its hadronic decay modes are being studied by BES Collaboration
at BEPC.
 
A lot of theoretical work have been done on charmonium and, in particular,
on these pseudoscalar charmonium 
decays[4,9-14]. Nonrelativistic quark model gives   
$\Gamma (\eta_c\rightarrow 2\gamma)=8.5 keV$ (using the observed 
$J/\psi$ leptonic width as input), while the QCD sum rule approach
predicts a value of $4.6\pm 0.4 keV$[12].
 
Recently, there have been significant progresses in the study 
of heavy quarkonium decays based on a more fundamental approach
of the NRQCD (nonrelativistic QCD) effective theory[15,16]. 
The factorization theorem was further discussed, and some important
issues (e.g., the infrared divergences in the P-wave state decay rates) 
were clarified in this study. The NRQCD theory combined with
nonperturbative lattice
simulations have achieved many interesting results on heavy quarkonium 
spectrum and decays[17,18,19].  

In this paper, we will use 
the Bethe-Salpeter (BS) formalism[20] to derive the decay amplitudes 
and to calculate the decay widths of heavy quarkonium.
The meson will be treated as a bound
state consist of a pair of constituent quark and antiquark 
(i.e., higher Fock states such as $|Q\bar Qg>$ and $|Q\bar Qgg>$ are
neglected, which may be justified to the
first order relativistic corrections of $S$ wave heavy quarkonium decays) 
and described by the BS wavefunction which
satisfies the BS equation.  
A phenomenological QCD-inspired interquark
potential will be used to solve for the wavefunctions and to 
calculate the decay widths.
Both relativistic and QCD radiative corrections
to next-to-leading order will be considered based on the factorization
assumption for the long distance and short distance effects. 

We first consider  the $\eta_c \rightarrow 2\gamma$ decay. 
This process proceeds 
via the $c\overline c$ annihilation. In the Bethe-Salpeter (BS) 
formalism the annilination matrix element
can be written as follows
\begin{equation}
\langle 0\mid \overline{Q}I Q\mid P\rangle =\int d^4qTr\left[ 
I(q,P)\chi _P(q)\right], 
\end{equation}
where $\mid P\rangle$ represents the heavy quarkonium state, 
$P (q)$ is the total (relative) momentum of the $Q\bar Q$, $\chi_P(q)$
is its four dimensional BS wave function, and
where $I(q,P)$ is the interaction vertex of the $Q\bar Q$ with other
fields (e.g., the photons or gluons ) which, in general, may also depend
on the variable $q^0$ (the time-component of the relative momentum). 
If $I(q,P)$ is independent of 
$q^0$ (e.g., if quarks are on their mass-shells in the annihilation), 
this equation can be written as
\begin{equation}
\langle 0\mid \overline{Q}I Q\mid P\rangle 
=\int d^3qTr[I 
(\stackrel{\rightharpoonup }{q},P)\Phi_{P}(\stackrel{\rightharpoonup }{q})],
\end{equation}
where 
\begin{equation}
\Phi_{P}(\stackrel{\rightharpoonup }{q})
=\int dq^0 \chi_P(q)
\end{equation}
is the three dimensional BS wave function of the $Q\bar Q$ meson.
Note that in this approximation 
the decay amplitude is greatly simplified and only the three dimensional
BS wave function is needed (but this does not necessarily require
the interquark interaction to be instantaneous).
In the BS formalism in the meson rest frame,
where ${\vec p_1}=-{\vec p_2}={\vec q},~P=(M,0)$, and  $p_1(p_2)$ is
the quark(antiquark) momentum, $M$ is the meson mass, we have 
\begin{eqnarray}
&&\Phi _{P}^{0^{-}}( \stackrel{%
\rightharpoonup }{q}) =\Lambda _{+}^1({\vec q})\gamma ^0( 1+\gamma ^0
) \gamma _5\gamma ^0\Lambda _{-
}^2(-{\vec q})\varphi(\stackrel{\rightharpoonup }{q}),\nonumber \\ 
&&\Phi _{P}^{1^{-}}( \stackrel{%
\rightharpoonup }{q}) =\Lambda _{+}^1({\vec q})\gamma ^0( 1+\gamma ^0
) \rlap/e\gamma ^0\Lambda _{-}^2(-{\vec q})
f(\stackrel{\rightharpoonup }{q}),
\end{eqnarray}
where $\Phi _{P}^{0^{-}}( \stackrel{%
\rightharpoonup }{q})$, and
$\Phi _{P}^{1^{-}}( \stackrel{%
\rightharpoonup }{q})$ represent the three dimensional wave
functions of $0^-$ and $1^-$ mesons respectively, 
$\rlap/e=e_{\mu}\gamma^{\mu}$, $e_{\mu}$ is the polarization
vector of $1^-$ meson, $\varphi$ and $f$ are scalar functions which can
be obtained by solving the BS equation for $0^-$
and $1^-$ mesons, and
$\Lambda_{+} (\Lambda_{-})$ are the positive (negative) energy 
projector operators
\begin{eqnarray}
&&\Lambda^{1}_{+}(\vec q)=\Lambda_{+}({\vec p}_1)=\frac 1{2E}(E+\gamma ^0%
{\vec \gamma}\cdot{\vec p}_1+m\gamma ^0),\nonumber\\
&&\Lambda^{2}_{-}(-\vec q)=\Lambda_{-}({\vec p}_2)=\frac 1{2E}(E-\gamma ^0%
{\vec \gamma}\cdot{\vec p}_2-m\gamma ^0),\nonumber\\
&&E=\sqrt{{\vec q}^2+m^2}.
\end{eqnarray}

For process $\eta_c\rightarrow 2\gamma$ 
with the photon momenta and polarizations $q_1, \epsilon _1$ and 
$q_2, \epsilon_2$, the decay amplitude can be written as
\begin{equation}
T=\langle 0\mid \overline{c}\Gamma _{\mu\nu}(q)c\mid\eta _c\rangle
\epsilon_1^\mu (\lambda_1)\epsilon_2^\nu (\lambda_2)+\langle 0\mid 
\overline{c}\Gamma_{\mu\nu}^{\prime}(q)c\mid\eta _c\rangle\epsilon
_2^\mu (\lambda _2)\epsilon _1^\nu (\lambda _1),
\end{equation}
where
$p_1(p_2)$ is the charm quark(antiquark) momentum, 
$p=p_1-q_{1}$, $p^{\prime}=p_1-q_{2}$,
$m$ and $M$ represent the masses of $c$ quark and $\eta_c$ meson
respectively, and where
\begin{eqnarray}
\Gamma _{\mu\nu}(q)=\gamma _\mu\frac{e^2e_Q^2}
{{\hat p}-m}\gamma _\nu,~~~ 
\Gamma _{\mu\nu}^{\prime}(q)=
\gamma _\mu\frac{e^2e_Q^2}{{\hat p}^{\prime}-m}\gamma _\nu,
\end{eqnarray}
$e_Q=\frac 23$ for $Q=c$.
Since $p_1^0+p_2^0=M$, as usual we take[1,13]
\begin{equation}
p_1^0=p_2^0=\frac M2.
\end{equation} 
Thus,
$p^{0}=\frac 12 M-q_{1}^0=0$, 
$p^{\prime 0}=\frac 12 M-q_{2}^0=0$,
the amplitude $T$ becomes
independent of $q^0$. Employing Eqs.~(2) and (4),
we get
\begin{equation}
T=B\epsilon^{\rho\sigma\mu\nu}q_{1\rho}q_{2\sigma}\epsilon_{1^\mu}
(\lambda_1)\epsilon_{2^\nu} (\lambda_2)e^2e_Q^2-
B^{\prime}\epsilon^{\rho
\sigma\mu\nu}q_{1\rho}q_{2\sigma}\epsilon_{1^\nu} (\lambda_1)\epsilon
_{2^\mu} (\lambda _2)e^2e_Q^2,
\end{equation}
where
\begin{eqnarray}
\label{a6}
&&B=B^{\prime},\nonumber \\
&&B=i\frac{2m}M\int d\stackrel{\rightharpoonup }{q}
\frac{\sqrt{\stackrel{%
\rightharpoonup }{q}^2+m^2}+m}
{\left( \stackrel{\rightharpoonup }{q}%
^2+m^2\right) \left( \stackrel{\rightharpoonup }{q}^2
+\stackrel{%
\rightharpoonup }{q}_1^2+m^2-2\stackrel{\rightharpoonup }{q}
\cdot \stackrel{%
\rightharpoonup}{q_1}\right) }\varphi (\stackrel{\rightharpoonup}{q}).
\end{eqnarray}
 Using $q_1\cdot \epsilon_1=0$ and $q_2\cdot \epsilon
_2=0$, it is easy to get the decay width
\begin{equation}
\label{a7}
\Gamma (\eta_c\rightarrow 2\gamma)=
3M^3\pi\alpha^2 e_Q^4\left| B\right|^2,
\end{equation}
In the nonrelativistic (NR) limit ($M\simeq 2m$, 
${\stackrel{\rightharpoonup }{q}}^2\rightarrow 0$)
\begin{equation}
\label{a8}
\left| B\right| ^2=\frac 1{2m^5}\left| \psi (0)\right| ^2,
\end{equation}
where we have used the relation
\begin{equation}
\label{a9}
\int \varphi (\stackrel{\rightharpoonup }{q})d
\stackrel{\rightharpoonup }{q}%
=\frac{\sqrt{M}}2\psi (0),
\end{equation}
where $\psi (0)$ is the Schr\"odinger wave function at origin in 
coordinate space.
Substituting (12) into (11), we get 
\begin{equation}
\label{a10}
\Gamma^{NR}(\eta _c\rightarrow 2\gamma )=
12\pi \alpha ^2e_Q^4\left| \psi
(0)\right| ^2/m^2,
\end{equation}
where
$\Gamma^{NR}(\eta _c\rightarrow 2\gamma )$ represents
the decay width of $\eta_c\rightarrow2\gamma$ in the nonrelativistic 
limit, which is consistent with that given in ref.[1].
The QCD radiative correction to this process has been given in ref.[1].
Recently, in the framework of NRQCD the factorization formulas for the
long distance and short distance effects were found to involve a double
expansion in the quark relative velocity $v$ and in the QCD coupling constant
$\alpha_s$[15]. To the next to leading order in both $v^2$ and $\alpha_s$, 
as an approximation, we may
write
\begin{equation}
\Gamma (\eta_c\rightarrow 2\gamma)=
3M^3\pi\alpha^2 e_Q^4\left| B\right|^2 (1-\frac {3.4\alpha_s(m_c)}{\pi}),
\end{equation} 
where the strong coupling constant $\alpha_s(m_c)$ is defined
in the $\overline {MS}$ scheme (the modified minimal subtraction
scheme).
By expanding $B$ in (10) in terms
of $\stackrel{\rightharpoonup }{q}^2
/m^2$, to the next to leading order of $v^2$  we have
\begin{equation}
\label{a18}
B=\frac{16i}{M(M^2+4m^2)}\int d\stackrel{\rightharpoonup }{q}
\varphi (\stackrel{%
\rightharpoonup }{q})(1-\frac {11}{12}
\frac{\stackrel{\rightharpoonup }{q}^2}{m^2}).
\end{equation}
We see that the relativistic kinematic effect is to suppress the
$\eta_c\rightarrow 2\gamma$  decay width.

For comparison with the process $J/\psi\rightarrow e^+e^-$,
we also give the decay amplitude for the $Q\bar Q$ annihilate into
an electron with momentum $k_1$ and helicity
$r_1$ and a positron with momentum $k_2$ and helicity $r_2$. Here the 
interaction vertex $I(P,q)=-ie\gamma_{\mu}$, which is independent of $q^0$,
and the amplitude can be written as
\begin{equation}
\label{a11}
T= e^2 e_Q\langle 0\mid \overline{c}
\gamma_{\mu} c\mid J/\psi \rangle \overline{u}_{r_1}(k_1)
\gamma^{\mu} v_{r_2}(k_2)\frac 1{M^2}.
\end{equation}
Define the decay constant $f_V$ by
\begin{equation}
\label{a12}
f_VMe_\mu\equiv \langle 0\mid \overline{c}\gamma _\mu c\mid J/\psi \rangle 
=\int d\stackrel{%
\rightharpoonup }{q}Tr[\gamma _\mu 
\Phi _{\stackrel{\rightharpoonup }{P}}(%
\stackrel{\rightharpoonup }{q})],
\end{equation}
where $e_{\mu}$ is the polarization vector of $J/\psi$ meson.
Then with (4) we find
\begin{equation}
\label{a13}
f_V=\frac{2\sqrt{3}}M\int d\stackrel{\rightharpoonup }{q}
(\frac{m+E}E-\frac{%
\stackrel{\rightharpoonup }{q}^2}{3E^2})
f(\stackrel{\rightharpoonup }{q}),
\end{equation}
where $E=\sqrt{\stackrel{\rightharpoonup }{q}^2+m^2}$.
Summing over the polarizations of the final states and
averaging over that of the initial states, it is easy
to get the decay width
\begin{equation}
\label{a14}
\Gamma (J/\psi \rightarrow e^{+}e^{-})=
\frac 43\pi \alpha ^2e_Q^2f_V^2/M.
\end{equation}
In the nonrelativistic limit it is reduced to the well known result
\begin{equation}
\label{a15}
\Gamma ^{NR}(J/\psi \rightarrow e^{+}e^{-})=
16\pi \alpha ^2e_Q^2\left| \psi
(0)\right| ^2/M^2.
\end{equation}
Including also the $QCD$ radiative correction[1],
we will get 
\begin{equation}
\label{a14}
\Gamma (J/\psi \rightarrow e^{+}e^{-})=
\frac 43\pi \alpha ^2e_Q^2\frac {f_V^2}M (1-\frac {5.3\alpha_s(m_c)}\pi).
\end{equation}
To the next to leading order of $v^2$, $f_V$ is expressed as
\begin{equation}
f_V=-\frac{4\sqrt{3}}M\int d
\stackrel{\rightharpoonup }{q}f(\stackrel{%
\rightharpoonup }{q})(1-\frac 5{12}
\frac{\stackrel{\rightharpoonup }{q}^2}{%
m^2}).
\end{equation}
Again, the relativistic kinematic correction is to reduce the
leptonic decay width. Comparing the two photon width with the leptonic 
width, we get
\begin{equation}
\label{a16}
R\equiv \frac{\Gamma (\eta _c\rightarrow 2\gamma )}{\Gamma (J/\psi
\rightarrow e^{+}e^{-})}=\frac 94M_{\eta
_c}^3M_{J/\psi }e_Q^2\frac{\left| B
\right| ^2}{f_V^2}(1+1.96\frac{\alpha
_s(m_c)}\pi ),
\end{equation}
and in the nonrelativistic limit it becomes 
\begin{equation}
\label{a17}
R^{NR}\equiv \frac{\Gamma ^{NR}(\eta _c
\rightarrow 2\gamma )}{\Gamma
^{NR}(J/\psi \rightarrow e^{+}e^{-})}=\frac
43(1+1.96\frac{\alpha _s(m_c)}\pi ).
\end{equation}

In fact, there are two sources of relativistic
corrections :
1) the correction of relativistic kinematics which appears explicitly 
in the decay amplitudes; 2) the correction due to inter-quark
dynamics (e.g. the well known Breit-Fermi interactions), 
which mainly causes the correction to the bound state wave functions.
In general, due to the attractive spin-spin force induced by one gluon
exchange for the $0^-$ meson, the $\eta_c$ wave function at origin 
becomes larger than its nonrelativistic value, one might expect the
width of $\eta_c\rightarrow 2\gamma$ to be enhanced after taking
relativistic corrections into account. 
However, because the kinematic relativistic correction to the decay rates 
is in the opposite direction and can be even larger,  
the overall relativistic correction to the decay width of $\eta_c$ is found 
to be negative.

To calculate the decay widths,
we need to know the wavefunctions  $\varphi(\vec q)$ for the $0^-$
meson and  $f(\vec q)$ for the $1^-$ meson, 
which are determined mainly by the long 
distance interquark dynamics. In the absence of a deep understanding
for quark confinement at present, we will follow a phenomenological
approach by using QCD inspired interquark potentials including both
spin-independent and spin-dependent potentials, 
which are supported
by both lattice QCD calculations and heavy quark phenomenology,
as the interaction kernel in the BS equation. We begin with the 
bound state BS equation[20] in momentum space
\begin{equation}
(\rlap/{q}_1-m_1)\chi_P(q)(\rlap/{q}_2+m_2)
={i\over{2\pi}}\int d^4 k G(P,q-k)\chi_P(k),
\end{equation}
where $q_1$ and $q_2$ represent the momenta of quark and 
antiquark respectively,
$G(P,q-k)$ is the interaction kernel which dominates the interquark dynamics.
In solving Eq.(26), 
we will employ the instantaneous
approximation since for heavy quarks the interaction is dominated by
instantaneous potentials. 
Meanwhile, we will neglect negative energy projectors in the
quark propagators which are of even higher orders. 
We then get the reduced Salpeter equation[20]
for the three dimensional BS wavefunction $\Phi_{P}({\vec q})$ defined
in (6)
\begin{equation}
\Phi_{P}({\vec q})={1\over{P^0-E_1-E_2}}\Lambda_{+}^1
\gamma^0\int d^3 k G( P,{\vec q}-{\vec k})
\Phi_{P}({\vec k})\gamma^0\Lambda^2_{-},
\end{equation}
where $G(P,{\vec q}-{\vec k})$ represents the 
instantaneous potential. 

We employ the following
interquark potrntials including a long-ranged confinement potential
(Lorentz scalar) and a short-ranged one-gluon exchange potential
(Lorentz vector)[21]
\begin{eqnarray}
&&{V(r)}={V_S(r)+\gamma_{\mu}\otimes\gamma^{\mu} V_V(r)},\nonumber \\
&&{V_S(r)}={\lambda r\frac {(1-e^{-\alpha r})}{\alpha 
r}},\nonumber \\
&&{V_V(r)}=-{\frac 43}{\frac {\alpha_{s}(r)} r}e^{-\alpha r},
\end{eqnarray}
where the introduction of the factor $e^{-\alpha r}$ is to regulate 
the infrared divergence and also to incorporate 
the color screening 
effects of the dynamical light quark pairs on the $Q\bar Q$ 
linear confinement potential[22]. 
In momentum space the potentials become[21]
\begin{eqnarray}
&&G( \stackrel{\rightharpoonup }{p})=G_S( \stackrel{%
\rightharpoonup }{p}) +\gamma_{\mu}\otimes \gamma^{\mu}
G_V( \stackrel{\rightharpoonup 
}{p}),\nonumber \\ 
&&G_S( \stackrel{\rightharpoonup }{p})=-\frac \lambda \alpha
\delta ^3( \stackrel{\rightharpoonup }{p})+\frac \lambda {\pi
^2}\frac 1{( \stackrel{\rightharpoonup }{p}^2+\alpha ^2) 
^2},\nonumber \\
&&G_V( \stackrel{\rightharpoonup }{p})=-\frac 2{3\pi^2}
\frac {\alpha_{s}(\stackrel{\rightharpoonup 
}{p})}{\stackrel{\rightharpoonup }{p}^2+\alpha ^2},
\end{eqnarray}
where $\alpha_{s}(\stackrel{\rightharpoonup }{p})$ is the  
quark-gluon running 
coupling constant and is assumed to become a constant of $O(1)$ as 
${\stackrel{\rightharpoonup }{p}}^2\rightarrow 0$
\begin{equation}
\alpha _s( \stackrel{\rightharpoonup }{p}) =\frac{12\pi }{27}%
\frac 1{\ln ( a+{\stackrel{\rightharpoonup }{p}^2}/{\Lambda 
_{QCD}^2%
}) }.
\end{equation}
The constants $\lambda$, $\alpha$, $a$ and $\Lambda_{QCD}$ are 
the parameters 
that characterize the potential. 

Substituting (4) and (29)
into Eq.(27), one derives the equation for the $0^-$ meson
wavefunction $\varphi(\vec q)$ in the meson rest frame
\begin{eqnarray}
& &M\varphi_1(\vec q)=(E_{q1}+E_{q2})\varphi_1({\vec q})\nonumber \\
& &={1\over{4E_{q1}E_{q2}}}\{-(E_{q1}E_{q2}+m_1m_2
+{\vec q}^2)
\int d^3 k (G_S({\vec q}-{\vec k})-4G_V({\vec q}-{\vec k}))\varphi_1(\vec k)
\nonumber  \\
& &-(E_{q1}m_2+E_{q2}m_1)\int d^3k(G_S({\vec q}-{\vec k})+
2G_V({\vec q}-{\vec k})) 
\frac{m_1+m_2}{E_{k1}+E_{k2}}\varphi_1(\vec k)
\nonumber  \\
& &+(E_{q1}+E_{q2})\int d^3 k G_S({\vec q}-{\vec k})({\vec q}\cdot{\vec k})
\frac{m_1+m_2}{E_{k1}m_2+E_{k2}m_1}\varphi_1(\vec k)
\nonumber  \\
& &+(m_1-m_2)
\int d^3 k (G_S({\vec q}-{\vec k})+2G_V({\vec q}-{\vec k}))
({\vec q}\cdot{\vec k}) 
\frac{E_{k1}-E_{k2}}{E_{k1}m_2+E_{k2}m_1}\varphi_1(\vec k),
\end{eqnarray}
where $E_{qi}=\sqrt{{\vec q}^2+m_i^2},~
E_{ki}=\sqrt{{\vec k}^2+m_i^2},~(i=1,2)$, and
\begin{equation}
\varphi_1(\vec q)=\frac{(m_1+m_2+E_{q1}+E_{q2})(E_{q1}m_2+E_{q2}m_1)}
{4E_{q1}E_{q2}(m_1+m_2)}
\varphi(\vec q).
\end{equation}
The normalization condition 
$\int d^3q Tr\{\Phi^{\dagger}({\vec q})\Phi({\vec q})\}
=(2\pi)^{-3}2M$ for the BS wavefunction leads to[21]
\begin{equation}
\int d^3 q \frac{(m_1+E_{q1})(m_2+E_{q2})}{8E_{q1}E_{q2}}
|\varphi({\vec q})|^2=\frac{M}{(4\pi)^3}.
\end{equation}
For the $1^-$ meson we have
\begin{eqnarray}
& &Mf_1(\vec q)=(E_{q1}+E_{q2})f_1(\vec q) \nonumber \\        
& & -\frac {E_{q1}+m_1+E_{q2}+m_2}
{4E_{q1}E_{q2}[3(E_{q1}+m_1)(E_{q2}+m_2)+{\vec q}^2]}
\{(E_{q1}E_{q2}+m_1m_2+{\vec q}^2)\times \nonumber \\
& & \int d^3 k(G_S({\vec q}-{\vec k})-2G_V({\vec q}-{\vec k}))
\frac{3(E_{k1}+m_1)(E_{k2}+m_2)+{\vec k}^2}
{E_{k1}+E_{k2}+m_1+m_2}f_1(\vec k) \nonumber \\
& & -2{\vec q}^2\int d^3 k(G_S({\vec q}-{\vec k})-2G_V({\vec q}-{\vec k}))
\frac{E_{k2}m_1+E_{k1}m_2}{m_1+m_2}f_1(\vec k) \nonumber \\
& & +(E_{q1}m_2+E_{q2}m_1)\int d^3 kG_S({\vec q}-{\vec k})
\frac{3(E_{k1}+m_1)(E_{k2}+m_2)-{\vec k}^2}
{E_{k1}+E_{k2}+m_1+m_2}f_1(\vec k) \nonumber \\
& & -(m_1+m_2)\int d^3 k(G_S({\vec q}-{\vec k})+4G_V({\vec q}-{\vec k}))
({\vec q}\cdot{\vec k})f_1(\vec k) \nonumber \\
& & -2(E_{q1}-E_{q2})\int d^3 k(G_S({\vec q}-{\vec k})-2G_V({\vec q}-{\vec k}))
({\vec q}\cdot{\vec k})\frac{E_{k1}+m_1}
{E_{k1}+m_1+E_{k2}+m_2}f_1(\vec k) \nonumber \\
& & +\int d^3 k(4G_S({\vec q}-{\vec k})-8G_V({\vec q}-{\vec k}))
({\vec q}\cdot{\vec k})^2 
\frac{f_1(\vec k)}{E_{k1}+m_1+E_{k2}+m_2} \nonumber \\
& & -2(m_1-m_2)\int d^3 kG_S({\vec q}-{\vec k})
\frac{E_{k1}-E_{k2}}{m_1+m_2}({\vec q}\cdot{\vec k})f_1(\vec k) \nonumber \\
& & +(E_{q1}+3E_{q2})\int d^3 kG_S({\vec q}-{\vec k})
({\vec q}\cdot{\vec k})f_1(\vec k) \nonumber \\
& & -(6E_{q1}+2E_{q2})\int d^3 kG_V({\vec q}-{\vec k})
{\vec q}\cdot{\vec k}f_1(\vec k)\},
\end{eqnarray}
where
\begin{equation}
f_1(\vec q)=\frac{E_{q1}+m_1+E_{q2}+m_2}{4E_{q1}E_{q2}}f(\vec q).
\end{equation}
The normalization condition 
$\int d^3q Tr\{\Phi^{\dagger}({\vec q})\Phi({\vec q})\}
=(2\pi)^{-3}2M$ for the BS wavefunction leads to[21] 
\begin{equation}
\int d^3 q \frac{(m_1+E_{q1})(m_2+E_{q2})}{8E_{q1}E_{q2}}
|f({\vec q})|^2=\frac{M}{(4\pi)^3}.
\end{equation}
To the leading order in the nonrelativistic limit, Eqs.(31) and (34) 
are just the ordinary nonrelativistic Schrodinger equation with simply a 
spin-independent linear
plus Coulomb potential. To the first order of
$v^2$, Eqs.(31) and (34) become the well known Breit equations 
for the $0^-$ and $1^-$ mesons with both
spin-independent and spin-dependent potentials from
vector (one-gluon) exchange and scalar (confinement) exchange.  

For the heavy quarkonium $c\bar c$ and $b\bar b$ systems,
$m_1=m_2=m$, Eqs.(31) and (34) become much simpler.  
 By solving these equations
 we can find the wave functions for the $0^-$ and $1^-$ mesons.
 Here not only the ground state (1S) wave functions but also the first 
 radial excitation wave functions (2S) are obtained. They are shown
 in Fig.1 and Fig.2.

Substituting the obtained BS wave functions into (10), (15), and
(19), (22), respectively,
we then get the decay widths for both $0^-$ and $1^-$ charmonium
states. In the calculation following parameters have been chosen 
\begin{eqnarray}
&&m_c=1.5GeV,\ \ \lambda=0.23GeV^2,\ \ \Lambda_{QCD}=0.18GeV,\nonumber\\ 
&&\alpha=0.06GeV,\ \ a=e=2.7183. 
\end{eqnarray}
With these parameters the
$2S-1S$ spacing and $J/\psi-\eta_c$ splitting are required to fit the
data. We then get
\begin {eqnarray}
\label{a23}
\Gamma (\eta_c\rightarrow 2\gamma)=6.2 keV, ~~~~
\Gamma (\eta_c^\prime\rightarrow 2\gamma)=1.8 keV,\nonumber\\
\Gamma (J/\psi\rightarrow e^+e^-)=5.6 keV, ~~~~
\Gamma (\psi^\prime\rightarrow e^+e^-)=2.7 keV.
\end{eqnarray}
Our results are satisfactory, as compared with the 
Particle Data Group experimental values[5] 
$\Gamma (\eta_c\rightarrow 2\gamma)=7.0^{+2.0}_{-1.7} keV,~
\Gamma (J/\psi\rightarrow e^+e^-)=5.36\pm 0.29 keV,~
\Gamma (\psi^\prime\rightarrow e^+e^-)=2.14\pm 0.21 keV.$ 
Here in above calculations the value of $\alpha_s (m_c)$ in the 
QCD radiative correction factor in (15) and (22) is chosen to 
be 0.29 (refs.[3,14]), which is also
consistent with our determination from the ratio of $B(J/\psi
\rightarrow 3g)$ to $B(J/\psi\rightarrow e^+e^-)$ (see refs.[4,9]).

In order to see the sensitivity of the decay widths to the parameters
especially the charm quark mass, we have also used other two sets of 
parameters  
\begin{eqnarray}
&&m_c=1.4GeV,\ \ \lambda=0.24GeV^2;\nonumber\\ 
&&m_c=1.6GeV,\ \ \lambda=0.22GeV^2,
\end{eqnarray}
with other potential parameters 
($\Lambda_{QCD},~\alpha,~a$) unchanged, and found
\begin {equation}
\label{a23}
\Gamma (\eta_c\rightarrow 2\gamma)=7.0 (5.5) keV, ~~~~
\Gamma (\eta_c^\prime\rightarrow 2\gamma)=1.7 (1.5) keV
\end{equation}
for $m_c=1.4~(1.6)~GeV$, where the experimental value of
$\Gamma (J/\psi\rightarrow e^+e^-)=5.36keV$ (as input)  
and the calculated ratio $R$ in (24) are used to give predictions for
the pseudoscalar decay widths. 

We see that for smaller charm quark masses
$\Gamma (\eta_c\rightarrow 2\gamma)$ gets enhanced. This tendency is in
line with the QCD sum rule result[12]. Our estimate that 
$\Gamma (\eta_c\rightarrow 2\gamma)=(6-7) keV$ is consistent 
with the $CLEO$ data[6] 
$\Gamma (\eta_c\rightarrow 2\gamma)=
(5.9 _{-1.8}^{+2.1}\pm 1.9) keV$, and the $E760$ data[7] $7\pm 3 keV$,
and slightly smaller than the $L3$ data[8]
$(8.0\pm 2.3\pm 2.4) keV$.  
Our results for $\eta_c$ and $\eta_c^\prime$ 
distinguish them from the nonrelavistic values, which can
be obtained by using the ratio $R^{NR}$ (25) and the experimental values of 
$\Gamma [J/\psi(\psi^\prime)\rightarrow e^+e^-]$
\begin{equation}
\label{a24}
\Gamma^{NR} (\eta_c\rightarrow 2\gamma)=8.5 keV,~~~
\Gamma^{NR} (\eta_c^\prime\rightarrow 2\gamma)=3.4 keV.
\end{equation}
In particular, our prediction $\Gamma (\eta_c^\prime\rightarrow 2\gamma)
=2 keV$ is significantly smaller than its nonrelativistic value. 

We may further use these results to give an estimate for the total
widths of $\eta_c$ and $\eta_c^\prime$. Note the branching ratio
\begin{equation}
B(P\rightarrow 2\gamma)\approx\frac{\Gamma(P\rightarrow 2\gamma)}
{\Gamma(P\rightarrow 2g)}=\frac{9\alpha^2e_Q^4}{2\alpha_s^2}\times
(\frac{1-3.4\alpha_s/\pi}{1+4.8\alpha_s/\pi})
\end{equation}
is free of the relativistic correction. Using the calculated two gamma
widths $\Gamma(P\rightarrow 2\gamma)=6.2 (1.8)keV$ for $P=\eta_c
(\eta_c^\prime)$ and the strong coupling constant at the mass of the
charm quark $\alpha_s(m_c)=0.26-0.29$[3,9,14], we will get
\begin{eqnarray}
\Gamma_{tot}(\eta_c)=17-23~MeV, \nonumber\\
\Gamma_{tot}(\eta_c^\prime)=5.0-6.7~MeV.
\end{eqnarray}
This is the prediction for the total widths up to the next to leading
order of QCD radiative corrections, but higher order corrections may
further modify this result. With the present Particle Data Group values[5]
$\Gamma_{tot}(\eta_c)=10.3^{+3.8}_{-3.4}MeV$ and 
$\Gamma (\eta_c\rightarrow 2\gamma)=7.0^{+2.0}_{-1.7} keV$,
however, a value of $\alpha_s(m_c)\approx 0.20$ will be indicated,
which is significantly lower than expected from other experiments
and theoretical studies on the QCD scale parameter.
Therefore, it will be very interesting to see the accuracy of the experiment
or to take the higher order QCD radiative correction more seriously.

Moreover, for the $b\bar b$ states, with $m_b=4.9GeV$ and other
potential parameters unchanged, we find
\begin {eqnarray}
\label{a23}
\Gamma (\eta_b\rightarrow 2\gamma)=0.46 keV, ~~~~
\Gamma (\eta_b^\prime\rightarrow 2\gamma)=0.21 keV,\nonumber\\
\Gamma (\Upsilon\rightarrow e^+e^-)=1.36 keV, ~~~~
\Gamma (\Upsilon^\prime\rightarrow e^+e^-)=0.78 keV.
\end{eqnarray}  
Here $\alpha_s(m_b)=0.20$[3,9,14] is used in the QCD radiative correctons.
We see that the relativistic corrections become smaller for $b\bar b$
than for $c\bar c$ states.

We finally discuss the relation between our approach and the NRQCD theory.
In fact, our decay widths can be written in terms of the
standard Schr$\ddot{o}$dinger wavefunction (with
relativistic corrections) $\psi_{Sch}(\vec q)$, which is related 
to $\varphi(\vec q)$ (or $f(\vec q)$) through the normalization
condition (33) (or (36)) which leads to
\begin{eqnarray}
\psi_{Sch}(\vec q)={1\over{\sqrt{M}}}
(\frac{m+E}{E})\varphi(\vec q), \nonumber \\
(2\pi)^3\int d^3 q~ \psi^{\ast}_{Sch}(\vec q)\psi_{Sch}(\vec q)=1.
\end{eqnarray}
For the above discussed pseudoscalar(P) and vector(V) heavy quarkonium 
decays (see (10), (15), and (19), (22)) to the next 
to leading order in $v^2$ and $\alpha_s$ 
we then have
\begin{equation}
\Gamma(P\rightarrow 2\gamma)=\frac{192\pi {\alpha}^2 e_Q^4M^2}
{(M^2+4m^2)^2}(1-\frac{3.4\alpha_s(m_Q)}{\pi})
|\int d^3 q (1-\frac{2{\vec q}^2}{3m^2})\psi_{Sch}(\vec q)|^2,
\end{equation}
\begin{equation}
\Gamma(V\rightarrow e^+e^-)=\frac{16\pi {\alpha}^2 e_Q^2}
{M^2}(1-\frac{16\alpha_s(m_Q)}{3\pi})
|\int d^3 q (1-\frac{{\vec q}^2}{6m^2})\psi_{Sch}(\vec q)|^2.
\end{equation}
In above expressions $M$ is the mass of the meson. 
In previous calculations we have taken $M$ as their observed values
for $\eta_c$ and $J/\psi$. However, we may 
take the on-shell condition, which assumes 
the quark and antiquark to be on the mass shell (see (8)) 
\begin{equation}
q^0_1=q^0_2=M/2=E=\sqrt{m^2+{\vec q}^2},
\end{equation}
to replace the observed value 
of the meson mass $M$ 
then (46) and (47) will become
\begin{equation}
\Gamma(P\rightarrow 2\gamma)=\frac{12\pi {\alpha}^2 e_Q^4}{m^2}
(1-\frac{3.4\alpha_s(m_Q)}{\pi})
|\int d^3 q (1-\frac{2{\vec q}^2}{3m^2})\psi_{Sch}(\vec q)|^2,
\end{equation}
\begin{equation}
\Gamma(V\rightarrow e^+e^-)=\frac{4\pi {\alpha}^2 e_Q^2}{m^2}
(1-\frac{16\alpha_s(m_Q)}{3\pi})
|\int d^3 q (1-\frac{2{\vec q}^2}{3m^2})\psi_{Sch}(\vec q)|^2,
\end{equation}
It is easy to see that to the first order of $v^2$,
in coordinate space (49) and (50) can be expressed as
\begin{equation}
\Gamma(P\rightarrow 2\gamma)=\frac{3\alpha^2 e_Q^4}{m^2}
(1-\frac{3.4\alpha_s(m_Q)}{\pi})
[|R(0)|^2
+\frac{4}{3m^2}Re(R^{\ast}(0)\nabla^2 R(0))],
\end{equation}
\begin{equation}
\Gamma(V\rightarrow e^+e^-)=\frac{\alpha^2 e_Q^2}{m^2}
(1-\frac{16\alpha_s(m_Q)}{3\pi})
[|R(0)|^2
+\frac{4}{3m^2}Re(R^{\ast}(0)\nabla^2 R(0))],
\end{equation}
where $R(0)$ is the Schr$\ddot{o}$dinger radial wavefunction 
at the origin of the $P~(P=\eta_c, \eta_c^\prime)$ or 
$V~ (V=J/\psi,\psi^\prime)$ meson. These expressions are exactly 
the same as that given
in ref.[15] with the NRQCD effective theory, 
if we identify our bound state wavefunctions with their
regularized operator matrix elements, i.e.:
\begin{equation}
R(0)={\sqrt {\frac{2\pi}{3}}}
{\bf \epsilon}\cdot<0|\chi^{\dagger}{\bf \sigma}\psi|V>,
\end{equation}
\begin{equation}
\nabla^2R(0)=-{\sqrt {\frac{2\pi}{3}}}
{\bf \epsilon}\cdot
<0|\chi^{\dagger}{\bf \sigma}({-i\over 2}
\stackrel{\leftrightarrow}{\bf D})^2
\psi|V>[1+O(v^2/c^2)].
\end{equation}

In the NRQCD theory,
the expectation values of the quark operators are well defined[15]
and can be calculated
with lattice simulations, which is a more fundamental method for describing
nonperturbative dynamics than the quark potential model. 
In our approach the wavefunctions (and
their derivatives) are estimated on the basis of the QCD-inspired
potential model by solving the BS equation. Although this is not a
first principle theory
and it is difficult to control the systematic accuracy 
within the potential model, it may provide a rather useful estimate
of the decay rates, since not only the zeroth order spin-independent
potential but also the first order spin-dependent potential i.e. the
Breit-Fermi Hamiltonian, which stems from one gluon exchange and has 
a good theoretical and phenomenological basis, are considered in
the calculation, and different quark massess are also chosen to 
estimate the uncertainties in the calculation. 
In fact, the potentials are required to reproduce
the observed mass difference between $\eta_c$ and $J/\psi$ and the $J/\psi$
leptonic decay width, and then give predictions for the pseudoscalar
mesons. This may reduce the uncertainty in the calculation of 
pseudoscalar decay widths.
Nevertheless, for more 
reliable estimates we hope 
that these decay widths of heavy quarkonium
can be eventually calculated from more fundamental
theoretical methods, e.g., the lattice QCD simulations.
It will be
interesting to see the numerical results  
in the NRQCD approach and compare them with our results.

In summary, we have estimated the photonic widths and hadronic widths for
pseudoscalar heavy quarkonium states, and the leptonic widths for vector
heavy quarkonium states as well, by taking into account both relativistic
and QCD radiative corrections. The photonic widths of $\eta_c$ and
$\eta_c^\prime$ tend to take lower
values than the nonrelativistic result. 
We hope that experiments especially the E835 and BES experiments or 
experiments at the
Tau-Charm Factory in the future will be able to have more accurate
measurements on the branching ratios and the total widths for the
$\eta_c$ and $\eta_c^\prime$ particles. This will provide the basis 
for testing theoretical predictions.

\vspace{10mm}

One of us (K.T.C.) would like to thank 
Prof. Y.F.Gu and Prof. K.K.Seth for very useful conversations on the 
BES and E760-E835 physics programs especially regarding the study for
$\eta_c$ and $\eta_c^\prime$ particles.
This work was supported in part by the National Natural
Science Foundation of China, and the State Education
Commission of China.


\begin{thebibliography}{99}
\bibitem{} R. Barbieri, R. Gatto and R. K$\ddot{o}$gerler, Phys. Lett.
	   {\bf B60}, 183(1976);\\
	   R. Barbieri, R. Gatto and E. Remiddi, Phys. Lett. {\bf B61},
	   465(1976); \\
		 R. Barbieri {\it et al.,} Nucl. Phys. {\bf B154}, 535(1979);
		 Nucl. Phys. {\bf B192}, 61(1981).               
\bibitem{} P. B. Mackenzie and G. P. Lepage, Phys. Rev. lett. 
{\bf 47}, 1244(1981). 
\bibitem{} W. Kwong, P. B. Mackenzie, R. Rosenfeld, and J. L. Rosner,
Phys. Rev. {\bf D37}, 3210(1988). 
\bibitem{} K. T. Chao, H. W. Huang, J. H. Liu, Y. Q. Liu, and
J. Tang, in Proceedings of {\it the International
Conference on Quark Confinement and the Hadron Spectrum,} Como, Italy,
June 1994, edited by N. Brambilla and G. M. Prosperi (World Scientific,
Singapore) p.306.
\bibitem{} Particle Data Group, L. Montanet {\it et al.,} Phys. Rev. 
{\bf D50 (3-I)}, 1171(1994).            
\bibitem{} $CLEO$ Collaboration, Phys. Lett. B{\bf 243}(1990)169.
\bibitem{} $E760$ Collaboration, K.K.Seth, in AIP Conference Proceedings
	   334 on Few-Body Problems in Physics, Williamsburg, VA, May 1994,
		 edited by F.Gross, p.248. 
\bibitem{} $L3$ Collaboration, Phys. Lett. B{\bf 318}(1993)575.
\bibitem{} K.T.Chao, H.W.Huang, and Y.Q.Liu, PUTP-94-20,
hep-ph/9503201, to be published in Phys. Rev.{\bf D}.
\bibitem{} Y. B. Ding, D. H. Qin, K. T. Chao and L. Zhou, High Energy 
Physics and Nuclear Physics {\bf 15}, 584 (1991).
\bibitem{} S. Godfrey and N. Isgur, Phys. Rev.{\bf D32}, 189 (1985).
\bibitem{} L.J.Reinders, H.Ruberstein and Yazaki,
	   Phys. Rep. {\bf 127}(1985)1; Phys. Lett. B{\bf 113}(1982) 411; \\
	   R. Kirschner and A. Schiller, Z.Phys. C{\bf 16}(1982)141.
\bibitem{} W. Y. Keung and I. J. Muzinich, Phys. Rev. {\bf D27},
1518(1983).
\bibitem{} H. C. Chiang, J. Hufner and H. J. Pirner, Phys. Lett.B{\bf 324},
(1994) 482.
\bibitem{} G. T. Bodwin, E. Braaten, and G. P. Lepage, Phys. Rev. {\bf D51},
1125 (1995);\\
     E. Braaten, Northwestern University preprint NUHEP-TH-94-22.
\bibitem{} G. T. Bodwin, E. Braaten, and G. P. Lepage, Phys. Rev. {\bf D46},
R1914(1992);\\
		 W. E. Caswell and G. P. Lepage, Phys. Lett. {\bf 167B}, 437(1986). 
	   228 (1983).
\bibitem{} A. El. Khadra, G. Hockney, A. Kronfeld, and P. Mackenzie,
Phys. Rev. Lett. {\bf 69}, 729 (1992).
\bibitem{} G. P. Lepage and J. Sloan, Nucl. Phys. B (Proc. Suppl.)
{\bf 34}, 417 (1994);\\
C. T. H. Davies {\it et al.,} OHSTPY-HEP-T-94-013.
\bibitem{} C. T. H. Davies {\it et al.,} Phys. Rev. {\bf D50}, 6963 (1994).


\bibitem{} E. E. Salpeter and H. A. Bethe, Phys. Rev. {\bf 84}, 1232(1951);\\
	   E. E. Salpeter, Phys. Rev. {\bf 87}, 328(1952);\\
		 For a BS formalism of the heavy quarkonium decays and related 
		 problems
		 see: W. Buchm$\ddot{u}$ller and S. H. H. Tye, Phys. Rev.
	   {\bf D24}, 132 (1981);\\ 
		 For a BS description of mesons in the instantaneous approximation
		 see also: K. T. Chao, Scientia Sinica {\bf A26}, 1050 (1983).
\bibitem{} K. T. Chao and J. H. Liu, in {\it Proceedings of the Beijing
Workshop on Weak Interactions and CP Violation}, Beijing, 1989, edited
by T. Huang and D. D. Wu (World Scientific, Singapore, 1990) p. 109;\\
	   J. Tang, J. H. Liu, and K. T. Chao, 
		 Phys. Rev. {\bf D51}, 3501 (1995) and references therein;\\
		 J. H. Liu, Ph.D thesis, Peking University, 1993 (unpublished);\\
\bibitem{} E. Laermann {\it et al}, Phys. Lett. {\bf B173}, 437(1986);\\
	   K. D. Born, Phys. Rev. {\bf D 40}, 1653 (1989). 
See also W. Kwong, J. L. Rosner, and C. Quigg, Annu. Rev. Nucl. Part. Sci.
{\bf 37}, 325(1987) and references therein for a review of
heavy quarkonium phenomenology.


\end{thebibliography}
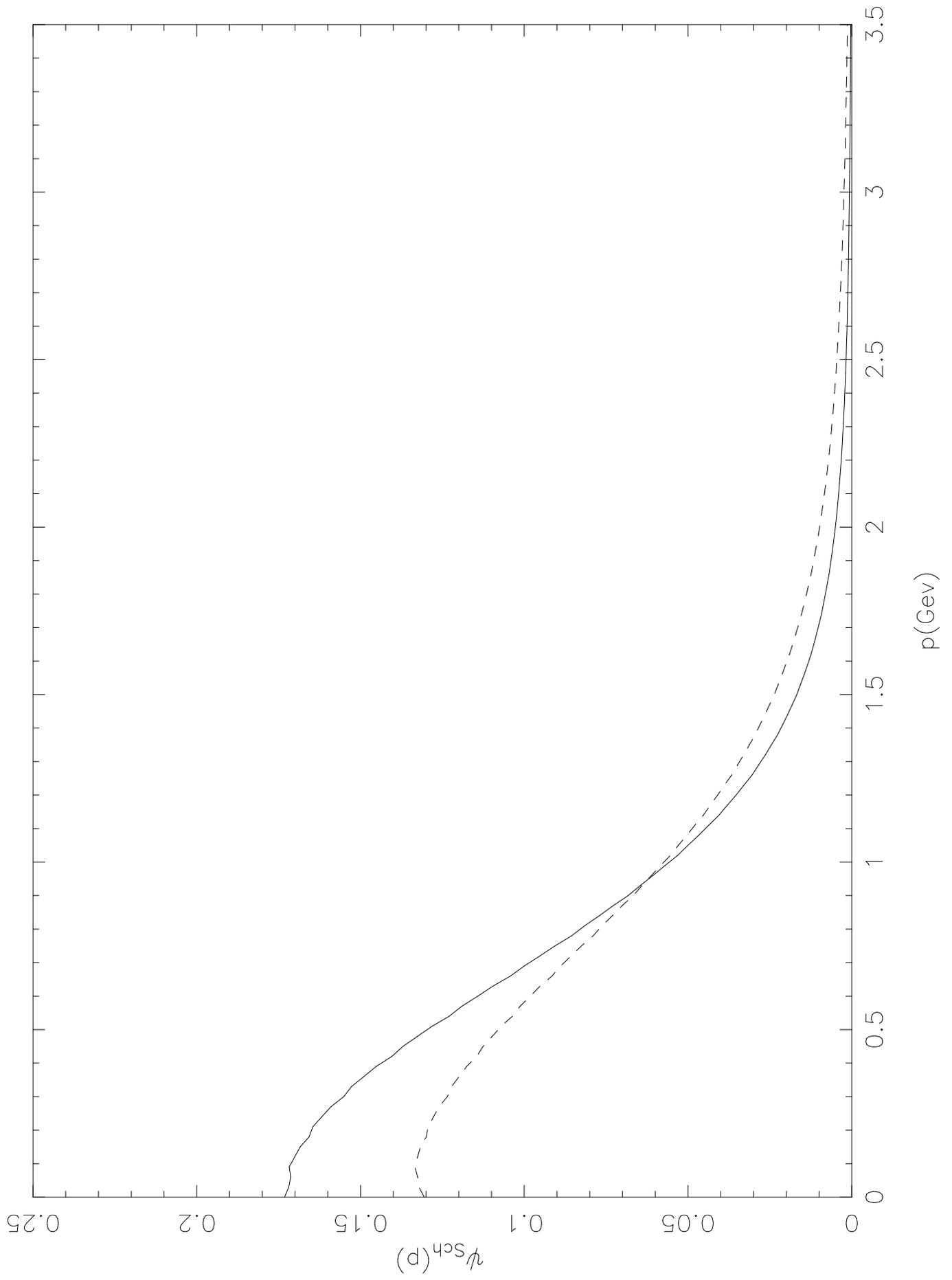
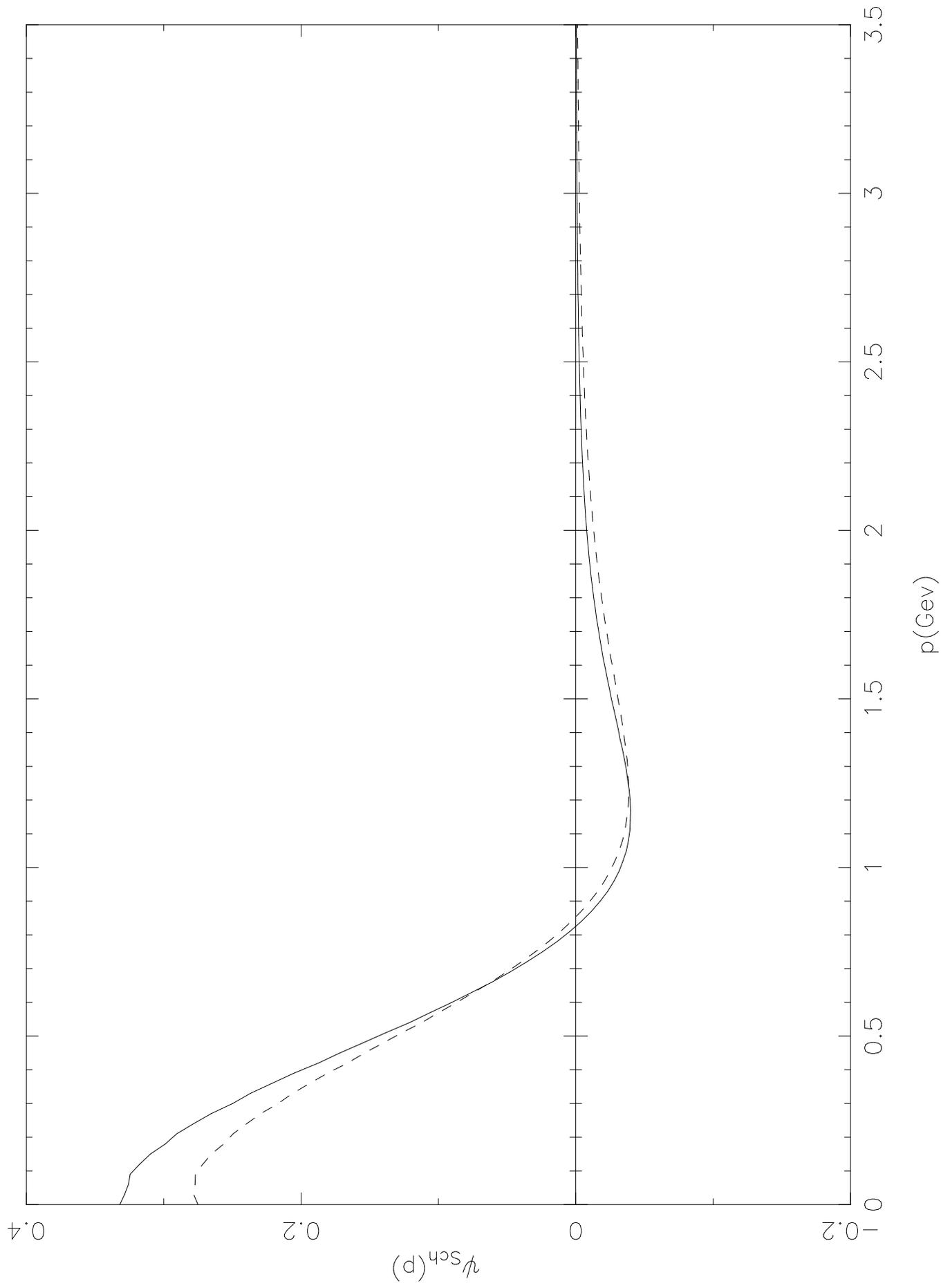
\end{document}